\documentclass[11pt]{article}
\newcommand{\BE}{\begin{equation}}
\newcommand{\EE}{\end{equation}}
\newcommand{\BA}{\begin{eqnarray}}
\newcommand{\EA}{\end{eqnarray}}
\newcommand{\Pp}{{\scriptstyle{{\rm P}}}}
\newcommand{\EC}{{\scriptscriptstyle{\rm EC}}}
\newcommand{\msbar}{\overline{\rm MS}}

\begin{document}

\thispagestyle{empty}

\vspace*{22mm}
\begin{center}
              {\LARGE\bf  The effective exponent $\gamma(Q)$ \\
              \vspace*{3mm} and the slope of the $\beta$ function}
\vspace{23mm}\\
{\large P. M. Stevenson}
\vspace{4mm}\\
{\it
T.W. Bonner Laboratory, Department of Physics and Astronomy,\\
Rice University, Houston, TX 77251, USA}

\vspace{30mm}

{\bf Abstract:}

\end{center}

\begin{quote}
The slope of the $\beta$ function at a fixed point is commonly thought to be 
RG invariant and to be the critical exponent $\gamma^*$ that governs the approach 
of any physical quantity ${\cal R}$ to its fixed-point limit: 
${\cal R}^*-{\cal R} \propto Q^{\gamma^*}$.  Ch\'{y}la has shown that this is not 
quite true.  Here we define a proper RG invariant, the ``effective exponent'' 
$\gamma(Q)$, whose fixed-point limit is the true $\gamma^*$. 
\end{quote}

\newpage

\setcounter{equation}{0}

     The $\beta$ function, $\beta(a) \equiv \mu \frac{d a}{d \mu}$, of a renormalizable 
quantum field theory is renormalization-scheme (RS) dependent.  The slope of the 
$\beta$ function at a fixed point, however, is commonly believed to be scheme invariant.  
That is not quite true.  

      The traditional argument \cite{Gross,Peterman} goes as follows.  Consider two RS's, 
primed and unprimed, whose renormalized coupling constants (couplants) are related by 
a general scheme transformation 
\BE
a'=a(1+v_1 a+ v_2 a^2 + \ldots) .
\EE
Their $\beta$ functions are related by 
\BE
\label{betatrans}
\beta'(a') = \frac{da'}{da} \beta(a).
\EE
If $\beta(a)$ vanishes at $a=a^*$ then $\beta'(a')$ will vanish at the corresponding 
$a'={a'}^*$.  (The scheme transformation could push ${a'}^*$ off to infinity, but let 
us assume that both $a^*$ and ${a'}^*$ exist and are finite.)  
The derivative of the $\beta$ function will transform as 
\BE
\label{term1trans}
\frac{d \beta'}{da'}=\frac{d\beta}{da} + \beta(a) \frac{d^2 a'}{da^2} \Big/ \frac{da'}{da}.
\EE
Since $\beta(a)$ vanishes at the fixed point, it would seem that 
\BE
{\mbox{    }}  {\mbox{    }}   {\mbox{    }}   {\mbox{    }}  {\mbox{    }}  {\mbox{    }}
{\mbox{    }}  {\mbox{    }}   {\mbox{    }}   {\mbox{    }}  {\mbox{    }}  {\mbox{    }}
{\mbox{    }}  {\mbox{    }}   {\mbox{    }}   {\mbox{    }}  {\mbox{    }}  {\mbox{    }}
{\mbox{    }}  {\mbox{    }}   {\mbox{    }}   {\mbox{    }}  {\mbox{    }}
\left. \frac{d \beta'}{da'} \right|_{*}= \left. \frac{d\beta}{da} \right|_{*}  
\quad\quad\quad {\mbox{\rm (not really true)}}.
\EE
Refs. \cite{Gross,Peterman} properly qualify this result with the proviso that 
$da'/da$ must not vanish and $d^2a'/da^2$ must not be singular, at $a=a^*$, so 
no criticism of these august authors is warranted.  Their unwary readers, however, may 
get the impression that these restrictions only refer to pathological or exceptionally 
rare RS transformations.   Ch\'{y}la \cite{chylafp} provides a salutary corrective to that 
attitude.  Indeed, a stark contradiction arises from trusting Eq.~(4), as we discuss in an 
appendix below.  
% We will attempt here to clarify matters 

Here we define the ``effective exponent'' $\gamma(Q)$, a $Q$-dependent ``scaling dimension'' 
associated with a specific physical quantity ${\cal R}$.  It is related to the slope of the $\beta$ 
function but has an extra term that is crucial for its RS invariance.  
    Our discussion will be at the 
formal level, except for some brief comments at the end.

     Consider some physical quantity ${\cal R}$, which may depend on several experimentally 
defined parameters.  We may always single out one such parameter, ``$Q$,'' with 
dimensions of energy, and make all other parameters dimensionless.  (The precise 
definition of $Q$ in any specific case is left to the reader.)  For definiteness we assume 
that the theory is asymptotically free as $Q \to \infty$, though our results are easily 
adaptable to the opposite case.  Also for definiteness we assume that ${\cal R}$ 
has a perturbation expansion
\BE
{\cal R}=a^\Pp (1+ r_1 a + \ldots),
\EE
although our key results apply whether or not ${\cal R}$ is calculated (or even 
calculable) perturbatively.
 
     Since ${\cal R}$ is a physical quantity and $Q$ is a physical parameter, the 
successive logarithmic derivatives of ${\cal R}$:
\BE
{\cal R}_{[n+1]} \equiv Q \frac{d {\cal R}_{[n]}}{dQ}
\EE
for $n=1,2,3,\ldots$, with ${\cal R}_{[1]} \equiv {\cal R}$, must be 
RS-invariant quantities, for any $Q$.  In particular, the combination
\BE
\label{gam1}
\gamma(Q) \equiv \frac{{\cal R}_{[3]}}{{\cal R}_{[2]}} 
\,\, = 1+Q \, \frac{d^2{\cal R}}{dQ^2} \Big/  \frac{d{\cal R}}{dQ}
\EE
is RS invariant.  It is the exponent of the local-power-law form of 
${\cal R}(Q)$ in the following sense:  Take the first three terms of the Taylor 
expansion of ${\cal R}$ about $Q=Q_0$ and fit them to the power-law form
\BE
{\cal R} \approx K + C Q^\gamma
\EE
to find
\BA
{\cal R}_0   \equiv \left. {\cal R} \right|_{Q=Q_0} & = & K + C Q_0^\gamma , \nonumber \\
{\cal R}_0'  \equiv \left. \frac{d{\cal R}}{dQ} \right|_{Q=Q_0} & = & \gamma C Q_0^{\gamma-1} , \\
{\cal R}_0''  \equiv \left. \frac{d^2 {\cal R}}{dQ^2} \right|_{Q=Q_0}  & =&
\gamma (\gamma-1) C Q_0^{\gamma-2} . \nonumber
\EA
These algebraic equations can be inverted to find the three parameters $K, C$, and 
$\gamma$.  (Note that $K$ is not ${\cal R}_0$ in general, though it is when $Q_0 \to 0$, 
assuming $\gamma >0$.)    
In particular, 
\BE
\gamma = 1+ Q_0 \frac{{\cal R}_0''}{{\cal R}_0'} ,
\EE
which is the $\gamma(Q_0)$ of Eq.~(\ref{gam1}).

   At high energies, where ${\cal R} \propto (1/\ln Q)^\Pp$, one has a negative   
$\gamma$, but as $Q$ is lowered $\gamma$ becomes positive.  As $Q \to 0$ it becomes the critical exponent 
$\gamma^*$ governing the approach of ${\cal R}$ to its fixed-point value 
${\cal R}^*$:
\BE
\label{gamstar}
({\cal R}^*-{\cal R}) \propto Q^{\gamma^*} \quad\quad {\mbox{\rm as}} \,\, Q \to 0.
\EE  
 
     In the perturbative expansion of ${\cal R}$, in some specific RS with renormalization 
scale $\mu$, the only $Q$ dependence resides in the series coefficients $r_i$.  For dimensional 
reasons, these can only depend on $Q$ through the ratio $Q/\mu$.  Thus, we have 
\BE 
\label{DA}
Q \frac{d{\cal R}}{dQ} = -\mu  \left. \frac{\partial {\cal R}}{\partial \mu} \right|_a ,
\EE
where the $\mu$ partial derivative is taken holding $a$ constant.  The total $\mu$ 
derivative of ${\cal R}$ vanishes:
\BE
\mu \frac{d {\cal R}}{d \mu}\,  =\,\, \mu \left. \frac{\partial {\cal R}}{\partial \mu} \right|_a 
+ \beta(a) \frac{d{\cal R}}{da} \, = \, 0 .
\EE
This RG equation says that  the $\mu$ dependence of the coefficients is cancelled by the 
$\mu$ dependence via the couplant $a$.   The two preceding equations lead to
\BE
\label{R2}
{\cal R}_{[2]} \equiv  Q \frac{d {\cal R}}{dQ} 
= \beta(a) \frac{d{\cal R}}{da}.
\EE
Since ${\cal R}_{[2]}$ is itself a physical quantity we can apply the same argument to it 
to get
\BE
\label{R3}
{\cal R}_{[3]} = \beta(a) \frac{d{\cal R}_{[2]}}{da} =
\beta(a) \left( \frac{d \beta}{da} \frac{d{\cal R}}{da} + 
\beta(a) \frac{d^2 {\cal R}}{d a^2} \right) .
\EE
Dividing Eq.~(\ref{R3}) by Eq.~(\ref{R2}) yields
our key result
\BE
\label{gam2}
\gamma(Q) =   \frac{d \beta}{da} + 
\beta(a) \frac{d^2 {\cal R}}{d a^2} \Big/ \frac{d{\cal R}}{d a} .
\EE

     [We digress briefly to recall a similar point made early in Ref.~\cite{KSS}.
The anomalous dimension of a Green's function ${\cal G}$ is conventionally 
defined as 
\BE
\gamma_{({\cal G})} \equiv \frac{\mu}{{\cal G}} \frac{d {\cal G}}{d \mu} 
= \frac{1}{{\cal G}} \left( \mu \left. \frac{\partial {\cal G}}{\partial \mu} \right|_a + 
\beta(a) \frac{d{\cal G}}{d a} \right) ,
\EE
which corresponds to the Callan-Symanzik equation \cite{CSym} for ${\cal G}$.  
It is {\it not} a physical quantity.  However, a physical quantity, an ``effective exponent'' 
for ${\cal G}$, can be defined as
\BE
\label{RcalG}
{\cal R}_{({\cal G})} \equiv - \left. \frac{\lambda}{{\cal G}} \frac{d}{d \lambda} 
{\cal G}(\lambda p_i,\mu,a(\mu)) \right|_{\lambda=1} . 
\EE                                                                                                                                                                                                                                                                                                                                                                                                                                                                                                        
(It could be written as $-\frac{Q}{{\cal G}} \frac{d{\cal G}}{dQ}$, given our convention
that $Q$ is the only dimensional physical parameter with all other parameters rendered 
dimensionless; e.g. $Q=p_1$ with the other parameters being $p_2/p_1, \ldots$.)  
The important point here is that the wavefunction-renormalization constant $Z_{({\cal G})}$ 
that multiplicatively renormalizes ${\cal G}$ is independent of the momentum arguments 
$p_i$ and cancels out in Eq.~(\ref{RcalG}).  
By the argument leading to Eq.~(\ref{DA}), we see that
\BE
\label{Geffexp}
{\cal R}_{({\cal G})} = \gamma_{({\cal G})} - \frac{\beta(a)}{{\cal G}} 
\frac{d {\cal G}}{da} ,
\EE
which is analogous to Eq.~(\ref{gam2}).]

    Returning to $\gamma(Q)$, it is instructive to check directly that Eq.~(\ref{gam2}) 
is invariant under scheme transformations.  The derivatives of ${\cal R}$ transform as 
\BA
\frac{d{\cal R}}{da'} \,\, & = & \frac{d{\cal R}}{da} \Big/ \frac{da'}{da} , \nonumber \\
\frac{d^2{\cal R}}{da'^2} & = & \frac{d}{da} 
\left( \frac{d{\cal R}}{da} \Big/ \frac{da'}{da} \right) \Big/ \frac{da'}{da} \\
     & = & \left( \frac{d^2{\cal R}}{da^2} - 
\frac{d{\cal R}}{da} \frac{d^2 a'}{da^2} \Big/ \frac{da'}{da} \right) 
\frac{1}{\left(\frac{da'}{da}\right)^2}.  \nonumber
\EA
Hence, the second term in Eq.~(\ref{gam2}) transforms as 
\BE
\label{term2trans}
\beta'(a') \frac{d^2 {\cal R}}{d a'^2} \Big/ {\frac{d{\cal R}}{d a'}} = 
\beta(a) \frac{d^2 {\cal R}}{d a^2} \Big/ {\frac{d{\cal R}}{d a}} -  
\beta(a) \frac{d^2 a'}{da^2} \Big/ \frac{da'}{da}.
\EE
Adding this to Eq.~(\ref{term1trans}) we see that 
\BE
\frac{d \beta'}{da'} + 
\beta'(a') \frac{d^2 {\cal R}}{d a'^2} \Big/ {\frac{d{\cal R}}{d a'}} 
\, = \,
\frac{d \beta}{da} + 
\beta(a) \frac{d^2 {\cal R}}{d a^2} \Big/ {\frac{d{\cal R}}{d a}} ,
\EE
confirming that $\gamma(Q)$ is genuinely scheme independent.  

    Further insight into $\gamma(Q)$ is the following observation.  Specialize to the 
case $\Pp=1$ (or define ${\cal R}_{\rm new} = {\cal R}_{\rm old}^{1/\Pp}$) 
and consider the ``effective charge'' (EC) renormalization scheme \cite{Grunberg} 
defined so that ${\cal R}=a(1+0+0+\ldots)$.  In this scheme 
$d^2{\cal R}/da^2=0$, so Eq.~(\ref{gam2}) reduces to 
\BE
\gamma(Q) = \frac{d \beta_{\EC}({\cal R})}{d {\cal R}}.
\EE
Thus $\gamma(Q)$, at any $Q$, is the slope of the EC $\beta$ function at 
the corresponding ${\cal R}$.  
In particular, in the infrared limit, the critical exponent $\gamma^*$ is the derivative 
of the EC $\beta$ function at the fixed point.   Moreover, from Eq.~(\ref{gam2}), we 
can say that $\gamma^*$ is the derivative of the $\beta$ function at the fixed point in 
any scheme for which $\frac{d{\cal R}}{d a}$ is non-zero and 
$ \frac{d^2 {\cal R}}{d a^2}$ is non-singular at $a=a^*$.   That 
includes a large class of possible RS's, but by no means is this ``almost all'' schemes 
\cite{chylafp}.  In general we must go back to Eq.~(\ref{gam2}) and carefully consider 
its infrared limit.  A similar point applies to Eq.~(\ref{Geffexp}).  For an instance where 
this subtlety arises see Ref.~\cite{explore}.

   An important open question concerns the ``universality,'' or otherwise, of $\gamma^*$.  
Is it the same for all perturbative physical quantities ${\cal R}$?  The question hinges on 
whether the EC couplants $a$ and $a'$ for two different physical quantities 
${\cal R}$ and ${\cal R}'$ always have $ \left. da'/da \right|_*$ non-zero and 
$ \left. d^2 a'/da^2 \right|_*$ non-singular.  Possibly yes, but it may well be that physical 
quantities segregate into distinct classes, each with a characteristic value of $\gamma^*$.  

    The preceding discussion has been entirely at the formal level.  In practice, 
of course, one uses some approximation to ${\cal R}$ and to $\beta(a)$.  A whole set 
of other issues then arises.  While physical quantities are scheme independent, perturbative 
approximations to them are not; scheme choice matters.  Fixed points can be made to 
appear or disappear under scheme transformations \cite{tH,KSS} when $\beta(a)$ and 
$\beta'(a')$ are each truncated and Eq.~(\ref{betatrans}) is satisfied only up to missing 
higher-order terms.  In the $\msbar$ scheme for QCD there is no fixed point at low $n_f$, 
but this may be entirely misleading.  In the EC scheme, or when the scheme choice is 
``optimized'' \cite{OPT}, one finds fixed-point behaviour for ${\cal R}_{e^+e^-}$ in 
both third \cite{lowen} and fourth \cite{OPTnew} order.  

    Other issues beyond the formal level are the related ones of perturbation-series 
divergence and power-suppressed non-perturbative terms, exponentially small in the couplant.  
 
   When approximating $\gamma(Q)$, or its infrared limit $\gamma^*$, the most 
meaningful result comes from its original definition, Eq.~(\ref{gam1}), with ${\cal R}$ 
replaced by its approximation.  For some schemes this is the same as using 
Eq.~(\ref{gam2}) with the ${\cal R}$ and the $\beta$ function replaced by their 
approximations, but in other schemes this may not be the case.

\vspace*{2mm}
{\bf Acknowledgment:}  I thank J. Ch\'{y}la for comments on the original manuscript.

\newpage

\section*{Appendix}

     We show here that a stark contradiction arises if the slope of the 
$\beta$ function at the fixed point,
\BE
\label{gammafalse}
\gamma_{\rm false}^* \equiv  \left. \frac{d\beta}{da} \right|_{*}, 
\EE
is taken to be a scheme-invariant quantity.   Writing the $\beta$ function as 
\BE
\label{betaexp}
\beta(a) = - b a^2 \sum_{i} c_i a^i 
\EE
(with $c_0 \equiv 1$ and $c_1 \equiv c$), the coefficients $b$ and $c$ are scheme invariant, 
but the $c_j$'s (for $j=2,3,\ldots$) are not.  These $c_j$'s, together with 
$\tau = b \ln (\mu/\tilde{\Lambda})$, can serve to parametrize the renormalization-scheme 
dependence \cite{OPT}.   (Since $\tau$ goes to $-\infty$ at the fixed point, it plays no role in 
our discussion here.)  Physical quantities are independent of the $c_j$'s (for $j=2,3,\ldots$), 
due to cancellation between the $c_j$ dependences of the perturbative coefficients and those of the 
couplant $a$.  The $c_j$ dependences of the couplant, $\partial a/ \partial c_j$ with $\tau$ 
and the other $c_i$'s held constant, are given by functions $\beta_j(a)$ defined in \cite{OPT}.  
In the fixed-point limit these tend to \cite{KSS}
\BE 
\label{astarcj}
\frac{\partial a^*}{\partial c_j} = \frac{b  {a^*}^{j+2}}{\gamma_{\rm false}^*}.  
\EE
This result follows easily by asking how the root $a^*$ of the equation 
$\sum_{i} c_i {a^*}^i  = 0$ changes as one specific $c_j$ is varied \cite{KSS}.  
Equivalently, if we define $B(a) \equiv \sum_{i} c_i {a}^i$, then $B^* \equiv B(a^*)$ is 
trivially RS invariant since it is zero in all schemes, so that
\BE
\left. \frac{\partial B^*}{\partial c_j} \right|_{a^*} + 
\frac{\partial a^*}{\partial c_j} \frac{d B^*}{d a^*} =0,
\EE
which leads directly to Eq.~(\ref{astarcj}).

  From Eq.~(\ref{gammafalse}) and $B^*=0$ we have 
\BE
\label{gamsum}
\gamma_{\rm false}^* = -b \sum_{i} i c_i {a^*}^{i+1} .
\EE
If $\gamma_{\rm false}^*$ were a physical quantity then we would have
\BE
 \left. \frac{\partial \gamma_{\rm false}^*}{\partial c_j} \right|_{a^*} + 
 \frac{\partial a^*}{\partial c_j} \frac{d \gamma_{\rm false}^*}{d a^*} =0.
 \EE
Using Eqs.~(\ref{gamsum}) and (\ref{astarcj}), and cancelling an 
overall $-b {a^*}^{j+1}$ factor, this would reduce to
\BE
j - \sum_{i} i(i+1) c_i {a^*}^i \Big/ \sum_{i} i c_i {a^*}^i =0.
\EE
But this equation would have to be true for all $j=2,3,\ldots$, which is clearly impossible since 
the second term is independent of $j$.

\end{document}